\documentstyle[twocolumn,prb,floats,aps]{revtex}
\preprint{Applied Physics Report 94-38}
\input psfig
\begin{document}
\draft
\title{
\vspace*{-10mm}
\begin{flushright}
{\large Applied Physics Report 94-38}
\end{flushright}
\vspace*{5mm}
\bf
Noise in a Quantum Point Contact due to a Fluctuating Impurity Configuration}

\author{J. P. Hessling\thanks{Electronic address:
hessling@fy.chalmers.se},$^{(1)}$ Yu. Galperin,$^{(2)}$
M. Jonson,$^{(1)}$ R. I. Shekhter,$^{(1)}$ and A.M. Zagoskin$^{(1)}$
}
\address{
$^{(1)}$Department of Applied Physics, Chalmers University of
Technology and G\"oteborg University,     S-412 96, G{\"o}teborg,
Sweden \\ $^{(2)}$Department of Physics, University of Oslo,
P.O. Box 1048 Blindern, N 0316 Oslo 3, Norway,
and A.F. Ioffe Physico-Technical Institute, 194021 St. Petersburg, Russia \\
\mbox{\rm \today} \\
\parbox{14cm}{\medskip \rm \indent
We propose a theoretical model for the low-frequency noise observed in a
quantum
point contact (QPC) electrostatically defined in the 2D  electron  gas  at  a
GaAs-AlGaAs interface. In such contacts electron scattering by soft 
impurity- or boundary potentials  coherently splits an incoming wave function
between different transverse modes.
Interference between these modes have been suggested to explain observed
non-linearities in the  QPC-conductance.
In this study we invoke the same mechanism and the time-dependent current due
to
soft dynamical impurity scattering in order to analyze the
%
low-frequency (telegraph-like)  noise  which  has  been  observed  along
with a nonlinear conductance.
For the simplified case of a channel
with two extended (current carrying) modes, a simple analytical formula for
the noise intensity is derived.
Generally we have found qualitative similarities between the noise and
the square of the transconductance. Nevertheless, incidentally there may
be situations when noise is suppressed but transconductance enhanced.
In comparison with the more traditional d.c. transport measurements
we believe that noise measurements can provide
additional information about the
dynamical  properties  of  QPCs.
}}
\maketitle
\narrowtext

\section{Introduction}
During the last several years problems related to electron transport through
ballistic point contacts have drawn a lot of attention  from  the  solid
state physics community. The most interesting feature of  the  so-called
quantum
point  contacts  (QPC)  is the nonlinear   character   of   their
current-voltage
($I-V$) characteristics.   This  nonlinearity  is
usually   explained   within  the  framework  of  conductance
quantization\cite{gh,kwh,zag1,pepper} (for a review see
Ref.~\onlinecite{BeenHout} and references therein).  According  to  this
concept the QPC forms a quantum channel that behaves as a wave guide for
electrons, the number of transverse modes being  dependent  both  on  gate
voltage
$V_g$ and source-drain voltage  $V_{sd}$.  Consequently, nonlinear  features
of
the $I-V_{sd}$  curve should be observed at driving voltages of the order of
the
mean spacing between the quantized transverse energy levels; i.e. for $eV_{sd}
\approx E_{\text{F}}/N$ [here $E_{\text{F}}$ is the Fermi energy and $N$ the
number
of modes]. For typical parameters this corresponds to
$V_{sd}  \approx  1$  mV. Nevertheless, nonlinear  structure in the
response has been  observed\cite{lin,lin2} at much lower voltages of order
$0.01$  mV.  One  could understand such a  behavior by taking  into  account a
resonant   structure in the
conductions steps (see e.g.  Refs.~\onlinecite{szs,kir,zagkul,xu}) due to
scattering caused by abrupt variations in shape at the entrance and exit
regions of the QPC.  An  explanation  appropriate  for QPCs with a smooth
geometry,
which probably corresponds to the experimental situation\cite{lin,lin2},  has
been  proposed  in   Ref.   \onlinecite{zsh1}. It is  based  upon the concept
of {\em coherent mode mixing} inside the contact; an  electron wave entering
the
contact  is  coherently split  between  different  transverse  modes due to a
scattering potential. During propagation along the channel the electron wave
packet
becomes deformed because the phase shifts gained  by  different  transverse
mode
components  are  not the same.  As a result, an  interference  structure in the
current appears because of mode mixing after a second scattering event.   Such
a
behavior is very similar to the well-known electrostatic Aharonov-Bohm  effect
(see  e.g.  Ref.~\onlinecite{BeenHout}). It is very important to notice that
the described model  does  not  rely  upon  a short range scattering potential
which can produce strong  backscattering.  On the contrary,  the effect
persists
even if the scattering potential is soft, which is the case in ballistic
structures
\cite{lbn,gj}.

The model introduced in Ref.~\onlinecite{zsh1} explains the qualitative
structure in the measured $I-V$ curves as well as their main quantitative
characteristics. We   believe   that  the  above  mentioned  mechanism  is
responsible  for the features observed in Refs.~\onlinecite{lin} and
\onlinecite{lin2}.

Along  with a  nonlinear  oscillating  contribution  to the $I-V$  curve of
QPCs,
one has also found a telegraph-like  low frequency noise\cite{lin3}.
Our aim here is to analyze noise properties of a QPC within the framework of
the
model introduced in Ref.~\onlinecite{zsh1}. Comparison of results on telegraph
noise
with available  experimental  data  could  support (or falsify) the above
mentioned
model.

The  paper  is  organized  as follows. In Section \ref{model} the model
employed  will   be   discussed.   General   analytical   expressions   for
telegraph-like  noise, as well as results of numerical calculations,
will  be  given  in Section \ref{noise}. In the last
Section \ref{discussion} we discuss main results and present our conclusions.

\section{The model } \label{model}

Following Ref.~\onlinecite{zsh1}, we model our QPC --- formed in a 2D electron
gas by gate electrodes --- as an adiabatically   smooth    channel  \cite{glks}
connecting  two equilibrium reservoirs. We assume
that the QPC
contains two scatterers. The origin of these scattering
centers, both taken to be static for the moment, can be the soft
potentials formed by one or several of the impurity atoms normally present
in QPCs because of doping. The gate electrodes are assumed to
provide a hard wall confining potential in the transverse direction, hence
creating a channel whose width is furthermore assumed to vary smoothly in the
longitudinal direction. The WKB approximation for the electron wave function is
therefore applicable. One finds\cite{glks}
\begin{equation} \label{WKB-WF}
\Psi(x,y,E)  = \sum_{n,\pm} a_{n}^{\pm}\chi_{n}^{\pm}(x,y,E)
\end{equation}
where
\begin{eqnarray} \nonumber
\chi_{n}^{\pm}(x,y,E)
& = &\sqrt{k_{n,\parallel}(E,\mp \infty) / k_{m,\parallel}(E,x)} \phi_{n,x}(y)
\\
& \times & \exp\left[i\int_{\mp \infty}^x dx^\prime
k_{n,\parallel}(E,x^\prime)\right], \label{not} \\
k_{n,\parallel}(E,x)
& = &k_{\text F} \sqrt{\varepsilon-\varepsilon_{n,\perp}(x) - ug(x)-vs(x)}.
\label{kl}
\end{eqnarray}
Here $E\equiv \varepsilon E_{\text F}$ is the total energy of the electron ---
$E_{\text F}$ being  the Fermi
energy in the leads at zero bias voltage, $V_{sd}=0$ --- while
$k_{\text F}=\sqrt{2mE_{\text F}/\hbar^2}$ is the Fermi wave vector and
$k_{n,\parallel}(E,x)$ the longitudinal wave vector along the channel.
The transverse
part of the wave function, $\phi_{n}(y)$, depends parametrically on the
longitudinal coordinate $x$; the corresponding `transverse' energy eigenvalue
is
$\varepsilon_{n,\perp}$ and is measured in units of the Fermi energy $E_{\text
F}$.
Hence
\begin{equation}
\varepsilon_{n,\perp}= \left(\frac{\pi n }{k_{\text F} d(x)}  \right)^{2},
\end{equation}
where  $d(x)$  is the coordinate dependent  width of the  channel.
The distribution of the electrostatic
potential caused by the applied  source-drain voltage is affected by
all charges within the contact region; it is described by the dimensionless
parameters $v= eV_{sd}/2E_{\text F}$  and $s(x)$ which appear in the
combination
$vs(x)$ in (\ref{kl}). In order to match the Fermi levels in the leads we must
obviously require that $s(\pm \infty)= \pm 1$.
The distribution along the channel for the gate voltage
modeled by the dimensionless parameter $u$, is described by the
dimensionless function $g(x)$. The plus sign ($+$) in (\ref{not}) corresponds
to
transmission from the left- to the  right reservoir, while a minus sign ($-$)
corresponds to transmission in the opposite direction.
Without loosing anything essential, 
we simplify our model by letting
$g(x)=1, \ v =0$ in expression (\ref{kl}) for the WKB wavefunctions.

The  interference  effects  of  interest to us  are  due to coherent
scattering by two scatterers. The first one splits
an incoming mode into several other modes which then
propagate   independently.  Making  use  of  the  unitarity of the scattering
matrix one can  show\cite{zsh1,bc}  that  such a
splitting  does  not  change  the  current  at  all  in  the   absence   of
backscattering (if the potential is soft enough).
The unitarity condition is simply a statement of conservation of
probability, total incoming flux must equal total outgoing flux.
The  second  scatterer makes an additional coherent mode mixing that
leads to an interference pattern in the total transmission (or  reflection)
coefficient, and in the current.

Let us assume that an electron enters from the left, is transmitted
after scattering against the left impurity ($L$),
propagates through the contact, passes the right impurity ($R$) after a
second scattering event,
and finally escapes the channel on the right hand side.
To describe the coherent splitting of the WKB wave function
(\ref{WKB-WF}) we introduce --- following Ref.~\onlinecite{zsh1} ---
for each scatterer a
unitary $(2N \times 2N)$
scattering  matrix,
\begin{equation}  \label{S}
\hat{S}\equiv \left( \begin{array}{cc}
{\hat r}^{-} & {\hat t}^{+} \\ {\hat t}^{-} & {\hat r}^{+}
\end{array} \right)
\end{equation}
such that
\begin{equation} \label{S2}
\left( \begin{array}{c}
{\bf \Psi}_{\text {out}}^+ \\
{\bf \Psi}_{\text {out}}^-
\end{array} \right) = {\hat S}
\left( \begin{array}{c}
{\bf \Psi}_{\text {in}}^- \\
{\bf \Psi}_{\text {in}}^+
\end{array} \right)
\end{equation}
The $(N \times N)$ submatrices $\hat{r}^{\pm}$ and $\hat{t}^{\pm}$
define the reflection from, and the transmission through the scatterer
of the components of the incident wave function. The plus/minus component
${\bf \Psi}_{\text {in}}^\pm$ refers
to the incident wave coming from the left/right reservoir,
while the  plus/minus component ${\bf \Psi}_{\text {out}}^\pm$ refers to
the outgoing wave from the scatterer propagating to the right/left.
The electron propagation right/left between the scattering events
is described by
diagonal {\em phase gain matrices} $\hat U$,
\begin{equation}  \label{U}
{\hat U}^{\pm}_{ij}  =  \delta_{ij} \exp \left[ \pm i \sigma_{j}(E)\right]
\end{equation}
where
\begin{equation} \label{Phase}
\sigma_{j}(E)  =  \int_{x_{L}}^{x_{R}} dx^\prime \,
k_{j, \parallel}(E, x^\prime)
\end{equation}
is the phase gained by the $j$-th mode between scatterers $R$ and $L$.

The general expression for the current is \cite{zsh1},
\begin{eqnarray}
I(V_{sd}) & = & \frac{2e}{h} \int dE \,  \delta   n_{\text F}(E) \,
\text{Tr} \left[ {\hat T^+}(E)^{\dagger} {\hat T^+}(E) \right] \nonumber \\
\delta n_{\text F}(E) & \equiv &
n_{\text F}(E- E_{\text F} - eV_{sd}/2) \nonumber \\
& - & n_{\text F}(E- E_{\text F} + eV_{sd}/2)
\end{eqnarray}
where $\delta n_{\text F}$ is  the  difference  between  the
Fermi-Dirac  distribution  functions  in the
reservoirs, and
${\hat T^+}(E)$ the total transfer matrix (for transfer from left to right)
for a particle with energy $E$,
so that
${\bf \Psi}_{\text{out}}^{+} = {\hat T^+} {\bf \Psi}_{\text{in}}^+.$
To lowest order, ${\hat T^+} = {\hat t}^{+}_{R}U^+{\hat t}^{+}_{L}$.
Making  use  of  Eqs.~(\ref{S})   and   (\ref{U}),
one   can
show  \cite{zsh1}  that  the  total  current can be expressed as
a sum of  two  parts:  $I=I_{\text{diag}}+I_{\text{int}}$.  The  first
part
\begin{equation}  \label{diag}
I_{\text{diag}}=  \frac{e}{\pi \hbar} \int dE \, \delta n_{\text F}(E)
\sum_{j} R_{jj}(E) L_{jj} (E)
\end{equation}
originates from diagonal parts of the
matrices
${\hat R} \equiv {\hat t}_R^{+\dagger}{\hat t}_R^+$ and
${\hat L} \equiv {\hat t}_L^+{\hat t}_L^{+\dagger}$. The second contribution,
\begin{equation}  \label{int}
I_{\text{int}} =  \frac{2e}{h} \int dE \, \delta n_{\text F}(E)
\sum_{j<k} 2 \text{ Re}\left( R_{jk} L_{kj} e^{i \sigma_{jk}} \right)
\end{equation}
where  $\sigma_{jk}  \equiv   \sigma_{j} - \sigma_{k}$, contains all
interference effects.
In the absence of backscattering Eq.~(\ref{diag}) reduces to the usual
expression for the current through a ballistic constriction, which
exhibits the well known conductance quantization with gate voltage
\cite{gh}. $I_{\text{int}}$ is the contribution from
interference   between   different   modes.  The expressions (\ref{diag})  and
(\ref{int}) that give the total current  were analyzed  in
Ref.~\onlinecite{zsh1}. They  are the  starting  points for calculating
low-frequency noise.

\section{Low-frequency noise } \label{noise}

\subsection{General consideration} \label{generalnoise}

One can imagine several mechanisms which lead to  low-frequency noise ---
external,  such  as  low-frequency  fluctuations  of  gate and source-drain
voltages, and internal, such as spatial rearrangements of the scattering
potentials. In  the  present study we consider the latter mechanism, which we
believe to be the  simplest one.

It  is  well  known  that there is some disorder in the vicinity of
small devices, even if they are of high quality.
In any  disordered  system defects with internal degrees of freedom are
present.
Interactions with a thermal bath can induce transitions between the
corresponding quantum-mechanical states (see  Ref.~\onlinecite{gkk} for a
review).
Usually,  such defects  switch  between two states, leading to a telegraph-like
noise pattern. Dynamical defects of this kind have been observed in metal point
contacts and tunnel junctions by several
authors (see, e.g.
Refs.~\onlinecite{ralls,rogers,buhrman,zimmerman,golding,buhrman4}) and have
been
called ``elementary fluctuators'', or EFs for brevity.

The  microscopic structure of the EFs is not completely clear. One of the
possible sources  of
two-state  defects  is  disorder-induced  soft  atomic
vibrations.   For  low  excitation  energies  the  vibrations  are  strongly
anharmonic and can be described as an atom or group of atoms moving  in  an
effective  double-well  potential.  Such  entities  are  known  as
\cite{and,phil}
two-level tunneling systems (TLS). They are responsible for the low
temperature properties of glassy materials. The generalization of the
TLS-model  for  higher  excitation  energies  has  been worked out in
Ref.~\onlinecite{karp}.   Dynamical   defects   produce  elastic  (or
electric)
fields,  slowly   varying  in time. Conduction electrons are then scattered by
these fields. Another possible origin  of fluctuations in  the  scattering
potential is electron  hopping between adjacent sites in the doped region of
the
device. The motion of the EFs leads to a variation of the scattering potential
in
the 2DEG region.

A quantitative theory is not yet completely worked
out for any of the two mechanisms mentioned above. In order to describe the
main
physical picture  we  use   a  simplified model; i.e. we assume that one of the
scatterers  (the left one, say) will maintain its static character, while the
other
one (right) is allowed to hop between two different spatial positions.
The hopping scatterer will from now on be denoted the elementary fluctuator
(EF).
Its spatial position $x_R$ is a random quantity, which we describe as
\begin{equation}  
x_{R}(t)= x_{R}^{0} - \frac{l}{2} \xi(t).
\end{equation}
Here $l$ is the hopping distance; for simplicity,
the transverse coordinate has been neglected, leading to an effectively
one dimensional description.
 The random quantity $\xi(t)$ is jumping between the two values
$\pm 1$ at random times, thus describing a telegraph-like process.

The hopping is induced by an interaction with a thermal bath. The transition
rate  $\Gamma_+$  from  the  state  $\xi=+1$  to  $\xi=-1$  and the reverse
rate
$\Gamma_-$ are determined by the nature of hopping and by the interaction
between
the EF  and the thermal bath. From the detailed balance principle we have
\begin{equation}  \label{Delta}
\Gamma_- /\Gamma_+ = \exp (-\Delta /k_{\text{B}}T)
\end{equation}
where  $T$  is  the  temperature, and  $\Delta$ the energy difference
between the states of the EF. Consequently,
at high enough temperatures ($k_BT\gg\Delta$)
the hopping rates are almost equal, while
at low temperatures there will be a significant difference between
the two. The dependence of $\Gamma =\Gamma_+ +
\Gamma_-$ on $\Delta$ and $T$ is determined by the hopping mechanism (see
the discussion in Ref. \onlinecite{gkk}).
If the transitions are due to quantum mechanical tunneling $\Gamma  \propto
\Delta^k$  where  $k=3$ for EF-phonon interaction and $k=1$ for EF-electron
interaction. If transitions  are  induced  by  activation on the other hand,
$\Gamma \propto \exp (-W/k_{\text{B}T})$ where $W$ is some activation energy
\cite{gkk}.

We assume that the only effect of the hops made by the scatterer is a
variation in the phases $\sigma_j$, gained by  conducting modes. As has been
shown
in Ref.~\onlinecite{zsh1} the oscillating part of the current is most sensitive
to
the phase of the modes. Therefore, we expect that the simple model we use
contains
the most important mechanism for the influence on the current
from the EFs.

\subsection{Analytical expression for noise intensity}

Noise is usually characterized by the current-current correlation function
\begin{equation}  \label{ngen}
S(\tau) = \langle I(t+ \tau) I(t) \rangle_t - \langle I(t) \rangle_t^{2}
\end{equation}
or  by  its Fourier transform $S(\omega)$ with respect to time $\tau$.
The symbol $\langle \cdots \rangle_t$ means average over time $t$
which is (under stationary conditions) just the same as
an average over the random process $\xi (t)$ (ergodicity).
Because the diagonal part
of the current is time-independent, only the interference part
 enters this expression.
Making use of the expression (\ref{int}) for $I_{\text{int}}$ we obtain the
following expression for the current-current correlation function
\begin{eqnarray} \label{corf}
S(\tau) &=& \left( \frac{2e}{h} \right)^{2}
\int dE \, \delta n_{\text F}(E) \int dE ^\prime \,
\delta n_{\text F}(E ^\prime) \nonumber \\
& \times &\sum_{j\ne k,l\ne m}M_{jklm}(E,E^\prime) \Phi_{jklm} ,
\end{eqnarray}
where
\begin{eqnarray}
M_{jklm} &=&
R(E)_{jk} L(E)_{kj}  R(E^\prime)_{lm} L(E^\prime)_{ml} \nonumber \\
\Phi_{jklm}&=&
\langle e^{i \sigma_{lm}(t,E^\prime)+i \sigma_{jk}(t+\tau,E)}\rangle_t
\nonumber \\ &-&
\langle e^{i \sigma_{lm}(t,E')}\rangle_t
\langle e^{i \sigma_{jk}(t,E)}\rangle_t .   \label{Phi}
\end{eqnarray}
Assuming that the longitudinal wavevector $k_{j, \parallel}$ varies slowly
as a function of  position $x_{R}$ in the channel, one can approximate
the phase as
\begin{eqnarray}
\sigma_{j}(t,E) & \approx &
\sigma_{j}^{0}(E) + w_j(E) \xi (t), \nonumber \\
\quad w_j(E) & \equiv & k_{j, \parallel}(E,x_{R}^{0}) l/2.
\end{eqnarray}
The superscript $0$ indicates that the phase (\ref{Phase}), should be
evaluated for $x_{R}^{0}$. We then arrive at the following approximate form
of the function (\ref{Phi}),
\begin{equation}  \label{Phi1}
\Phi_{jklm} \approx
e^{i[ \sigma_{jk}^0(E) + \sigma_{lm}^0(E^\prime)]}
{\cal G}\left[w_{jk}(E),w_{lm}(E^\prime)|\tau\right]
\end{equation}
where $w_{ji} (E) \equiv w_j (E) -w_i(E)$, and
\begin{eqnarray}  \label{GenFun}
{\cal G}(x,y|\tau) & = & K(x,y|\tau)-K(x,y|\infty), \nonumber \\
K(x,y|\tau) & = & \left< e^{ix \xi (t+\tau) + iy \xi (t)} \right>_t.
\end{eqnarray}
Note   that  the  function  $K(x,y|\tau)$  is  known \cite{gard} as  the
{\em   generating function} for the random process $\xi (t)$.

Now, in order to evaluate $K(x,y|\tau)$ let us analyze the Master
equation for the conditional probability $Q( \xi^\prime, t^\prime| \xi, t)$
of finding the  value  $\xi  =  \xi^\prime$ at the time $t^\prime$ under the
condition $\xi(t)=\xi$. It reads  (cf. Ref. \onlinecite{gard}),
\begin{eqnarray} \label{Master1}
\frac{ \partial Q( \xi^\prime , t^\prime | \xi, t)}{ \partial t}
& - & \xi^\prime \Gamma_+ Q(+1, t^\prime | \xi, t) \nonumber \\
& + & \xi^\prime \Gamma_- Q(-1, t^\prime | \xi, t) =0,
\end{eqnarray}
with initial condition $Q( \xi^\prime, t | \xi,  t)=  \delta_{  \xi^\prime,
\xi}$. Taking into account the sum rule
$$Q(+1, t^\prime | \xi, t)+ Q(-1, t ^\prime | \xi, t) =  1,$$
we find the following solution to (\ref{Master1}),
\begin{eqnarray}
Q( \xi ^\prime , t^\prime | \xi, t)
& =&   \frac{1}{2\Gamma} \left[  \Gamma
 - \xi^\prime ( \Gamma_+ - \Gamma_-)
+ \xi^\prime e^{ -  \Gamma| t^\prime - t |} \right.
\nonumber \\
&\times& \left. \left(\Gamma \delta_{+1, \xi} -
\Gamma \delta_{-1, \xi} +  \Gamma_+ - \Gamma_-
\right)  \right].
\label{MsSol}
\end{eqnarray}

The expectation values of all odd and even powers
of $ \xi (t)$ and its products
can now readily be evaluated
(since $ \xi$ is only allowed to assume the values $ \pm 1$). One finds
\begin{eqnarray}
\label{MV1}
\langle \xi^{2k}(t^\prime) \xi^{2n} (t)\rangle
& = & 1, \nonumber \\
\langle \xi^{2k} (t^\prime)  \xi^{2n +1} (t) \rangle
& = & \langle \xi (t) \rangle = (\Gamma_- - \Gamma_+)/\Gamma, \nonumber \\
\langle \xi^{2k+1} (t^\prime)  \xi^{2n+1} (t)\rangle
& = & \langle
\xi (t^\prime) \xi (t)\rangle \equiv C (|t^\prime - t |) .
\label{MV2}
\end{eqnarray}
The   function  $C|(t^\prime-t|)$  defined in Eq.~(\ref{MV2}) can  be
calculated
using  its definition, which can be re-expressed as
\begin{equation}  \label{cf}
C(t^\prime,t)=\sum_{\xi,\xi^\prime = \pm 1}\xi^\prime \xi
Q(\xi^\prime, t^\prime|\xi, t) P(\xi, t).
\end{equation}
$P(\xi,  t)$ is here the one-event probability to find the value $\xi$ at
time $t$. We find
\begin{equation} \label{K_2}
C (t^\prime, t) = \frac {1}{\Gamma^2}\left[
(\Gamma_+ - \Gamma_-)^2
 + 4 \Gamma_+ \Gamma_-
e^{ -\Gamma |t^\prime - t|} \right].
\end{equation}
Expanding the exponential in the generating function
(\ref{GenFun}) and making use of the expectation values
(\ref{MV1}), together with
(\ref{K_2}) we  obtain,
\begin{eqnarray} \label{FinGenFun}
K(x,y|\tau)&=&\cos x \cos y \nonumber \\
 &-& \sin x \sin y \left[
\frac{ (\Gamma_+ - \Gamma_-)^2 }{ \Gamma^2 }
+ \frac{ 4 \Gamma_+ \Gamma_- }{ \Gamma^2 }
e^{ - \Gamma |\tau|  } \right] \nonumber \\
 &-& i \frac{ \Gamma_+ - \Gamma_- }{ \Gamma } \sin ( x + y ).
\end{eqnarray}
For the difference function ${\cal G}(x,y|\tau)$ defined in (\ref{GenFun}) we
get
\begin{eqnarray}  \label{G1}
G(x,y|\tau)&=&-\sin  x  \sin  y\frac {4\Gamma_+ \Gamma_-}{\Gamma^2}e^{-\Gamma
|\tau|}
\nonumber \\
&=&  -\frac  {\sin  x  \sin  y}{\cosh^2(\Delta/2k_{\text{B}}T)}e^{-\Gamma
|\tau|}.
\end{eqnarray}
Here  we  have  used  the relation  (\ref{Delta})  between  $\Gamma_+$ and
$\Gamma_-$. We see that the quantity
${\cal G}\left[w_{jk}(E),w_{lm}(E^\prime)|\tau\right]$  in     Eq.
(\ref{Phi1}) can be factorized into functions of $E$ and
$E^\prime$.  By their mere construction, the matrices $\hat R$ and $\hat L$
are Hermitian. Hence one can reduce the double sum as follows,
\begin{eqnarray}
&&\sum_{j\ne k} \sin w_{jk}(E) R(E)_{j,k} L(E)_{kj} e^{i \sigma_{jk}^{0}(E)}
\nonumber \\ && =
2 i \sum_{j<k} \sin w_{jk}(E) \, \text{Im}  \left[
R_{jk}(E) L_{kj}(E) e^{i \sigma_{jk}^{0}(E)} \right].
\end{eqnarray}
Taking the Fourier transform with respect to $\tau$ we find the noise
spectrum to be a Lorentzian since the current-current correlation function
is exponentially decreasing in time.

Collecting the partial results above, we obtain a general expression for the
noise of the form
\begin{eqnarray} \label{sp1}
S( \omega )  &=&
\frac{1}{\cosh^2(\Delta/2k_{\text{B}}T)} {\cal L} (\omega)  \Lambda^2,
\nonumber \\
{\cal L}(\omega)& =&
\frac{1}{ \pi } \frac{\Gamma }{ \Gamma^{2} + \omega ^{2} },
\end{eqnarray}
where
\begin{eqnarray}
&&\Lambda  = \frac{2e}{\pi \hbar} \int dE \, \delta n_{F}(E)
\nonumber \\
&&\times
\sum_{j<k}
\sin w_{jk}(E,x_R^0)
\, \text{ Im}  \left[R_{jk}(E) L_{kj}(E) e^{i \sigma_{j,k}^{0}(E)}
\right]. \label{sp1lambda}
\end{eqnarray}
The expression (\ref{sp1}) is a product of three factors. The first one,
$\cosh^{-2}  (\Delta /2k_{\text{B}}T)$, makes it clear that
a telegraph-like  noise  can only  appear if the  EF  can  hop  due  to
interactions  with  a  thermal  bath.  The typical excitation
energy is of order $k_{\text{B}}T$. If $\Delta  \gg  k_{\text{B}}T$  the  EF
cannot  be
excited  and  will  remain in its lowest energy state forever. The
second  factor  is  a simple Lorentzian function with  characteristic  width
$\Gamma$.

The third factor in (\ref{sp1}), $\Lambda^2$, contains the same quantities as
expression (\ref{int}) for the interference part of the current.
It is of interest to see whether or not
the quantity $\Lambda$ can be related to the
measured  current-voltage  characteristics.  To  address  this
question we
consider the simplified case of a narrow QPC
allowing for only two propagating modes.
At low temperatures  the variation with energy of the matrices $\hat R$
and $\hat L$ and phases $\sigma$ in an energy interval of  order
$k_{\text{B}}T$  near  the  Fermi
energy  can  be neglected. The resulting expressions for the current and for
the noise spectrum are in this case very simple,
\begin{eqnarray} \label{sp2}
I_{\text{int}}&=&I_{\max} \cos \left[\sigma_{12}^{0}(E_F) + \gamma\right],
  \\
S(\omega)&=& I_{\max}^2\frac {\sin^2\left[w_{12}(E_F)\right]\sin^2
\left[\sigma_{12}(E_F) + \gamma\right]}{\cosh^2 (\Delta/2k_{\text{B}}T)}
{\cal L}(\omega),  \\
I_{\max} & = &  \frac{4 e }{h } \int dE \, \delta n_{F}(E) \left|
R_{12}(E)_{12} L_{21}(E)\right| \label{sp2last}
\end{eqnarray}
The phase difference $ \sigma_{12}^{0}(E_{\text F})$ is of the order of
$k_{\text F} L$, where $L$ is the spatial separation between the scatterers,
$ x_{L} - x_{R}^{0}$, and $\gamma$  some  phase  shift
of no interest here.  $I_{\max}$  is  the
maximum  value  of  the  interference  current with respect to variations in
the
external parameters. %
%
Qualitative conclusions about expressions (\ref{sp2})-(\ref{sp2last})
will be drawn in Section \ref{discussion}.
%
Next we will present a numerical analysis for the situation
with several modes. The  role  of the channel  geometry  will  also be
discussed.

\subsection{Numerical analysis}

In the case when there are more than two modes propagating through the
quantum point contact, one has to rely upon numerical simulations. We have
calculated the noise as a function of gate- and source-drain voltages. The
quantities usually measured experimentally are the current and the
transconductance, defined as \begin{equation}
\label{trans_cond}
G(V_{sd},V_{g})= 
\frac{\partial I}{\partial V_{g}}.
\end{equation}
Only the interference part of the  current has been considered  because  far
enough   from  the  conduction  steps  the  diagonal  part  is  practically
independent of both gate- and source-drain voltages.
The ranges  of all  parameters were limited  in such a way as to give a
constant number of propagating modes.

For a numerical evaluation of the general expression (\ref{sp1}) for the noise,
we need
a definite form of $R_{jk}(E) L_{kj}(E)$. This quantity is the product of the
two scattering probabilities for scattering from mode $j$ to $k$ by
the left scatterer and the analogous probability --- but time-reversed and with
exchanged mode  indices ---  for the right scatterer.
In a completely symmetric channel (including the
impurities) with  respect to
the coordinate along the channel,  $R_{jk}(E) L_{kj}(E) = |L_{kj}(E)|^2$ .
To simplify the calculations we assume that the only screening of the
impurity potential that exists originates from  electrons within
the gate  electrode. This screening can approximately be described by a proper
image charge. If the impurity and image charges are separated from the 2DEG by
distances $a$ and $b$, respectively, the effective potential is
\begin{equation} \label{pot}
V(r) = A_0\left[ \frac{1}{\sqrt{(r^2+a^2)}} -
\frac{1}{\sqrt{(r^2+b^2)}}  \right],
\end{equation}
where $\bf r$ is the in-plane coordinate, while $A_0$ contains
the necessary physical constants.

The scattering probability can be calculated from the Golden rule,
\begin{equation}\label{Lkj}
L_{kj}(E) = \frac{2 \pi}{ \hbar} \left|<f|V(x,y)|i> \right|^2 \delta(E_f-E_i)
\end{equation}
where $i$ and $f$ denote the initial and final states.

In the calculation of the matrix element above we have simplified the
geometry and regarded  the channel to be of constant width.
For $r \gg \max{(a,b)}$, we find that $V(r) \propto  1/r^3$. The
potential (\ref{pot}) is  hence short-ranged,
and its influence may be ignored for large $r$. In the opposite limit
of  small values of $r$,
we get $V(r) = A_0  (1/a -1/b)[1-(r/c)^2]$, where
 $c=a b / \sqrt{(a^2 + a b + b^2)/2}$.
The typical range of $V(r)$ is $c$.

Since the expansion for small $r$ of the Lorentzian function
$[(1+(r/c)^2]^{-1}$  has the same form as the expansion of
the actual potential,
and since the precise value of $V(r)$ for large $r$ is not
very crucial as long as
it decays  quickly enough, we may use the approximation
\begin{equation}
V(r) = A_0 \frac{b-a}{ab} \frac{1}{1+(r/c)^2}.
\end{equation}

The problem of finding $L_{kj}(E)$ can be further simplified if the
potential  factorizes, so that
$V(x,y) = V(x) V(y)$. To get a rough estimate, this factorization
can be done as,
\begin{equation}
V(x,y) \approx A_0 \frac{b-a}{ab}\frac{1}{1 + (x/c)^2)} \Theta[1-(y/c)^2]
\end{equation}
Here we have reduced the $y$ dependence to a step function,
but kept the smooth form in the $x$ direction. In this way we prevent
the  Fourier transform of the potential with respect to
$x$ from being oscillatory, which seems unphysical.

Finally, evaluating the matrix element (\ref{Lkj})
we arrive at the expression
\begin{equation}
L_{kj}(E) = A \frac{(b-a)^3k_F}{(b^3-a^3)k_{k,\parallel}}
 \frac{1-(-1)^{k-j}}{2(k-j)^2}
                 e^{ -2 \left|k_{k,\parallel} -  k_{j,\parallel} \right| c},
\end{equation}
where $A$ contains only fundamental physical constants including an
effective dielectric constant.
Due to the fact that the potential in the transverse direction is an
even function of $y$,  we only have a non zero scattering probability
between modes of the  same parity (i.e. when $k-j$ is an even number).

\begin{figure} \label{fig:1}
\centerline{\psfig{figure=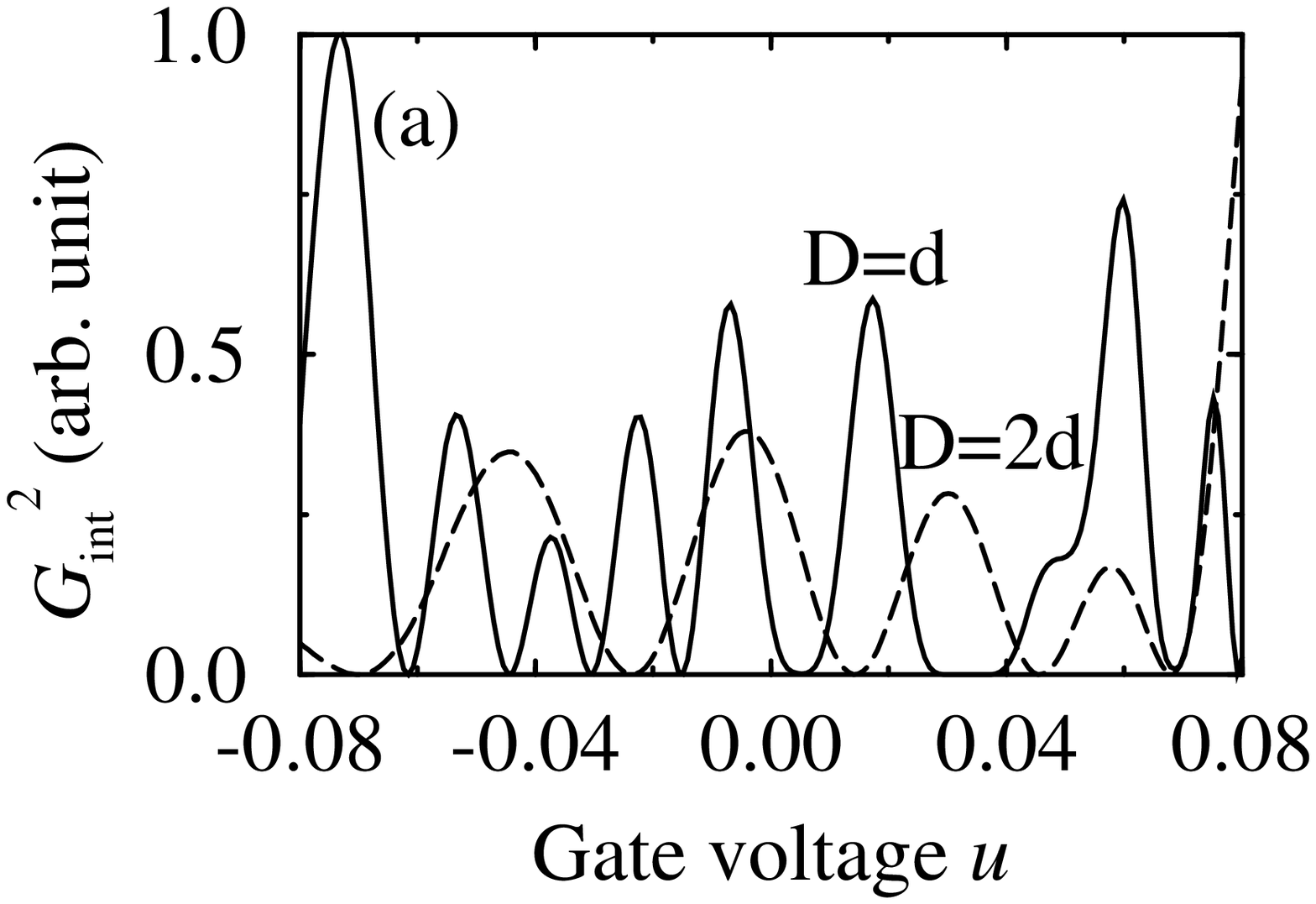,width=9cm}}
\vspace*{3 mm}
\centerline{\psfig{figure=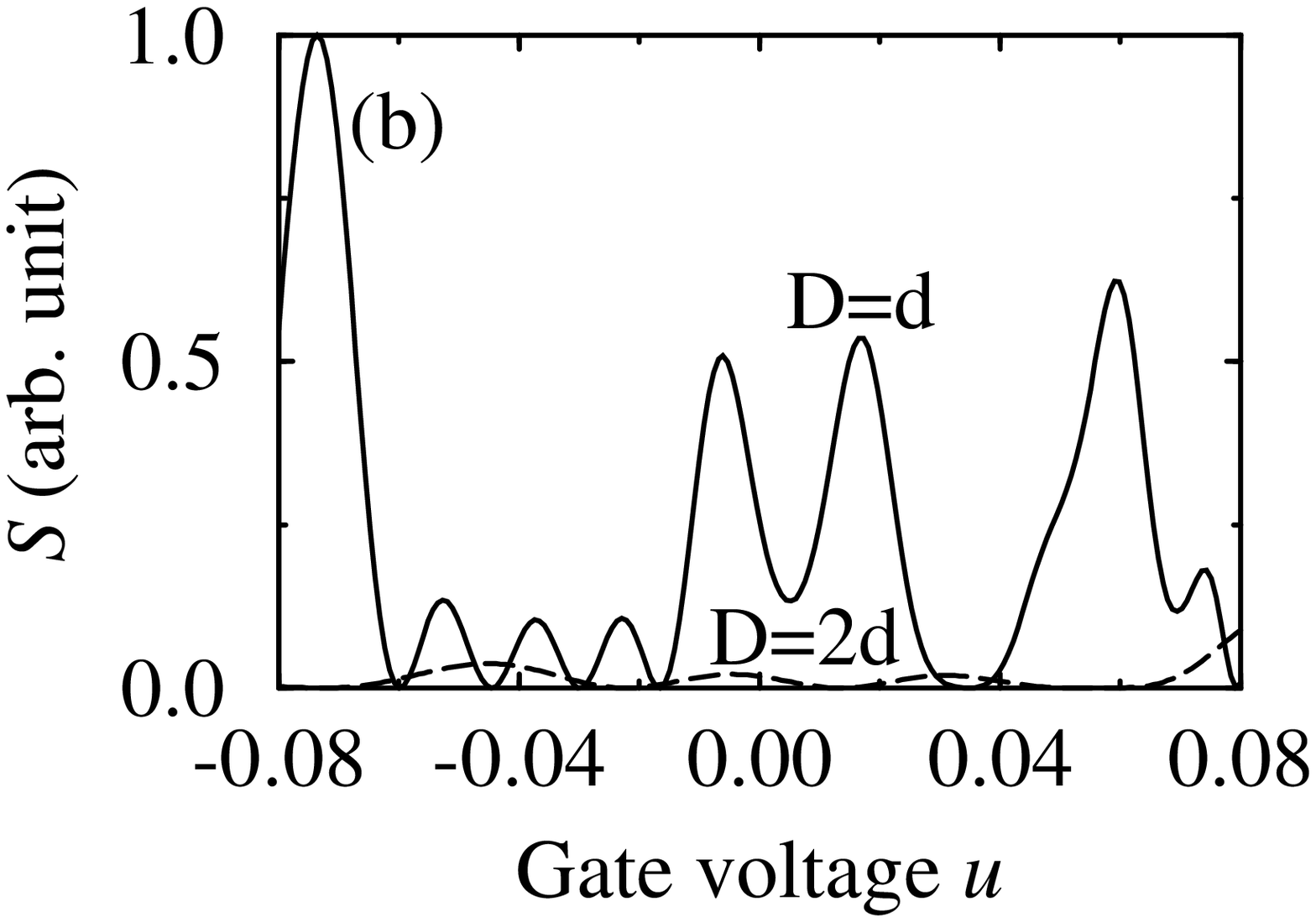,width=9cm}}
\vspace*{5 mm}
\caption{
Square of transconductance (panel $a$) and noise (panal $b$)
as functions
of dimensionless gate voltage parameter $u$ ($V_g= u E_{\text F}$).
Two different parabolic channels are used for which the relations between the
maximum ($D$) and minimum ($d)$ width are $D/d=1,2$.
The lengths of the channels are in all cases 
$79.6\lambda_{\text F}$,
where $\lambda_{\text F}$ is the Fermi wavelength.
The minimum width is chosen so as
to let 5 modes propagate through the contact.
The hopping distance $l$ for the
fluctuator was taken to be 
$0.796\lambda_{\text F}$.
The calculation was done for the source-drain voltage
$V_{sd}= 0.02 E_{\text F}$
and for the temperature
$T=10^{-3}T_{\text F}$, $E_{\text F}$ and $T_{\text F}$
being the Fermi energy and temperature, respectively.}
\end{figure}

In Fig.~1 (panel $a$)
the square of transconductance is plotted versus the dimensionless gate voltage
$u =eV_g/E_{\text F}$.
The  QPC  is  assumed to have a parabolic shape, the scatterers being placed
symmetrically with respect to the narrowest point. Hence, the width of the
channel is
\begin{eqnarray}
d(x)&=& d + (D-d) \left(
\frac{x-\bar x}{L/2}\right)^{2},
\nonumber \\
 {\bar x} &=&(x_{R}^{0}+x_{L})/2.
\end{eqnarray}
We have chosen the minimum width $d$ so that five modes are allowed to
propagate
through the contact, $k_{\text F} L = 500$, and $k_{\text F} l = 1$. The latter
choice needs a digression. The typical displacement $l$ of an EF in the course
of
thermal-bath-induced fluctuations depends strongly on
the microscopic nature of the EF.
If the EF is produced by structural disorder in the
vicinity of the channel, one can expect $l$ to be of the order of
an interatomic distance, and $k_{\text F} l \ll 1$.
If the EF originates from electron hopping in the doped region
of the structure, it is reasonable to expect that
$l$ is of the order of the average distance
between the impurities in the doped region.
In this case the product $k_{\text F} l$ can be as large as $1-10$
depending on size and shape of the structure.
Another point to note is that a mechanical displacement of the scatterer
is not the only reason for variations in the scattering phase $k_{\text F} l$.
Rearrangements in an extended defect containing
several atoms can also cause variations in the
scattering phase of order unity.
Our phenomenological model does not allow for
explicit calculations of the scattering phase, so we keep the simplest
description of the mechanical displacement of the
scatterer and use a reasonable value for the product $k_{\text F} l$.

Taking the gate voltage parameter to be $u=0$ corresponds to
considering the center
of the plateaus in the $I-V_g$ curve. Our calculations were made for two
different values of the ratio between the channel parameters,
$D/d=1$ and $2$. The noise intensity is shown in Fig.~1 (panel $b$).
As is clearly seen, there is a strong qualitative correlation between
the noise and the square of the transconductance.

{}From our expressions (\ref{sp2}) for noise
and current, the similarity between the graphs seems plausible.
Indeed, both quantities have oscillating parts of the form
$\sin^{2} \left(\sigma_{1,2}^{0} + \gamma \right)$.
In addition, the expression for the noise contains another
more slowly varying phase
factor, $\sin^2\left[w_{12}(E_F)\right]$,
causing some modulation of the noise relative to the square of
transconductance.
The noise is weak in the regions $-0.06<u<-0.02$ for $D/d=1$
simply because this second phase factor
causes almost every term to be small.
For the case $D/d=2$ the phase itself is so small that the noise almost
vanishes.
Since there are many interference terms
in the sum, some pairs of channels might lead to a stronger suppression than
others.
%
%
\begin{figure} \label{fig:3}
\centerline{\psfig{figure=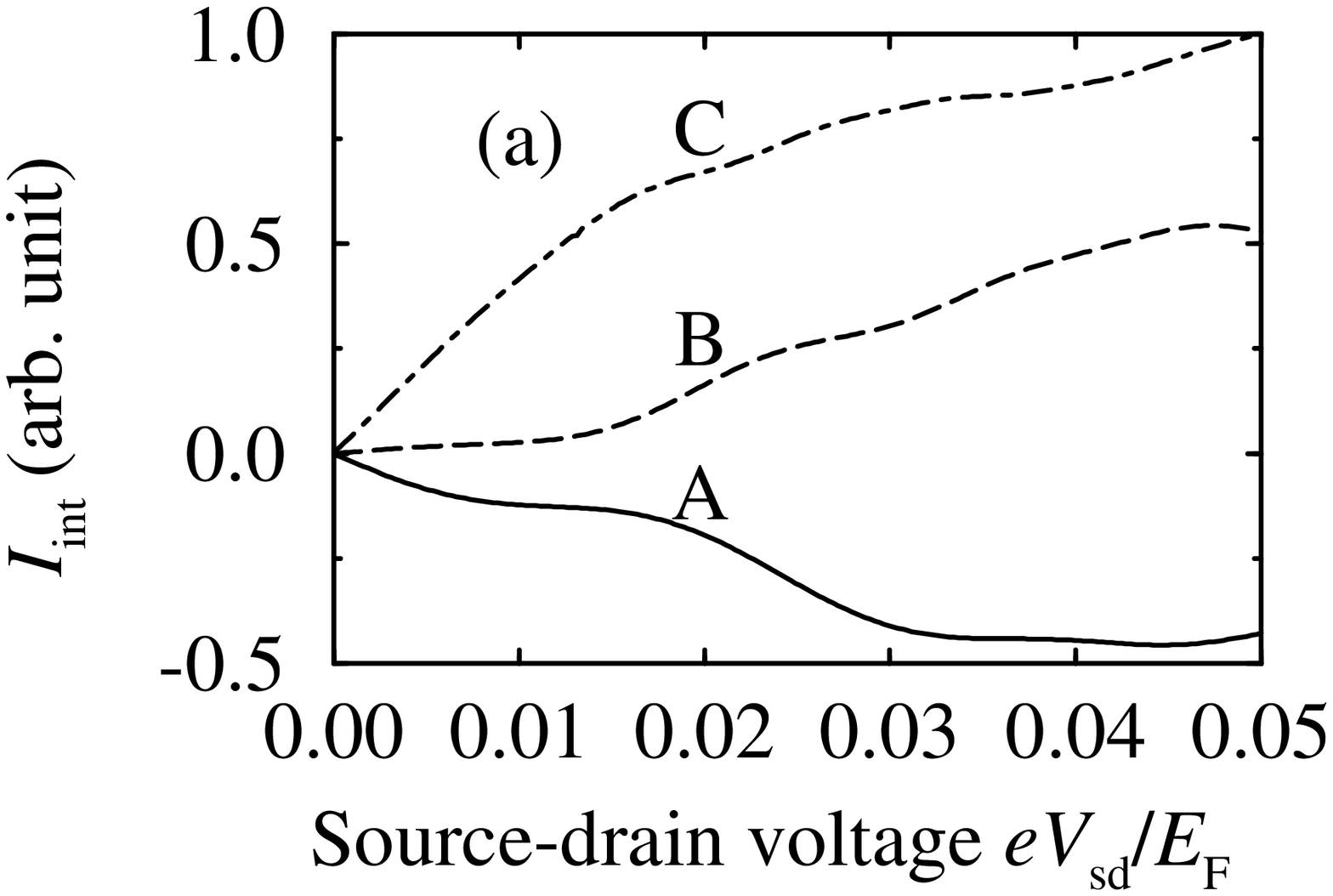,width=9cm}}
\vspace*{3 mm}
\centerline{\psfig{figure=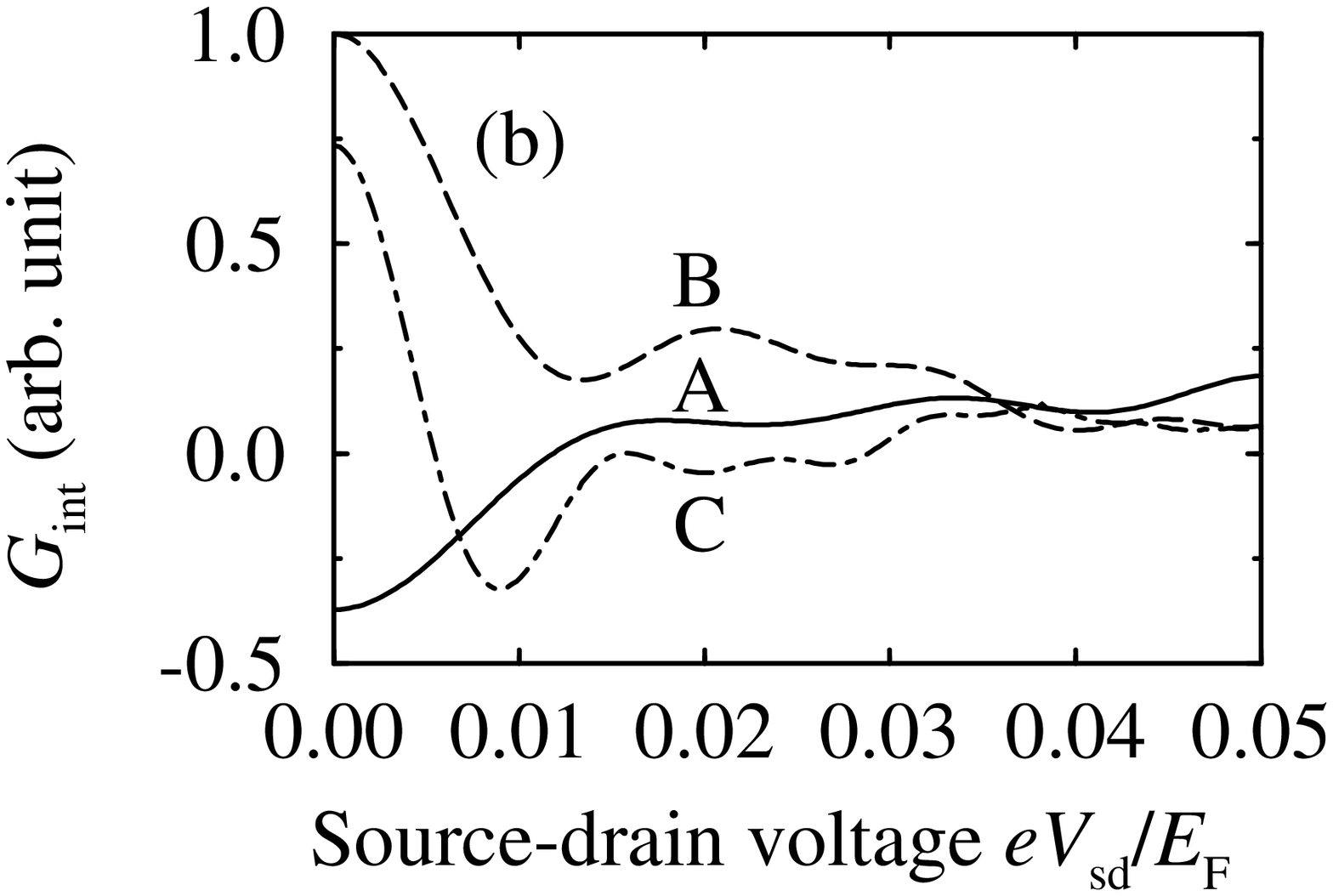,width=9cm}}
\vspace*{3 mm}
\centerline{\psfig{figure=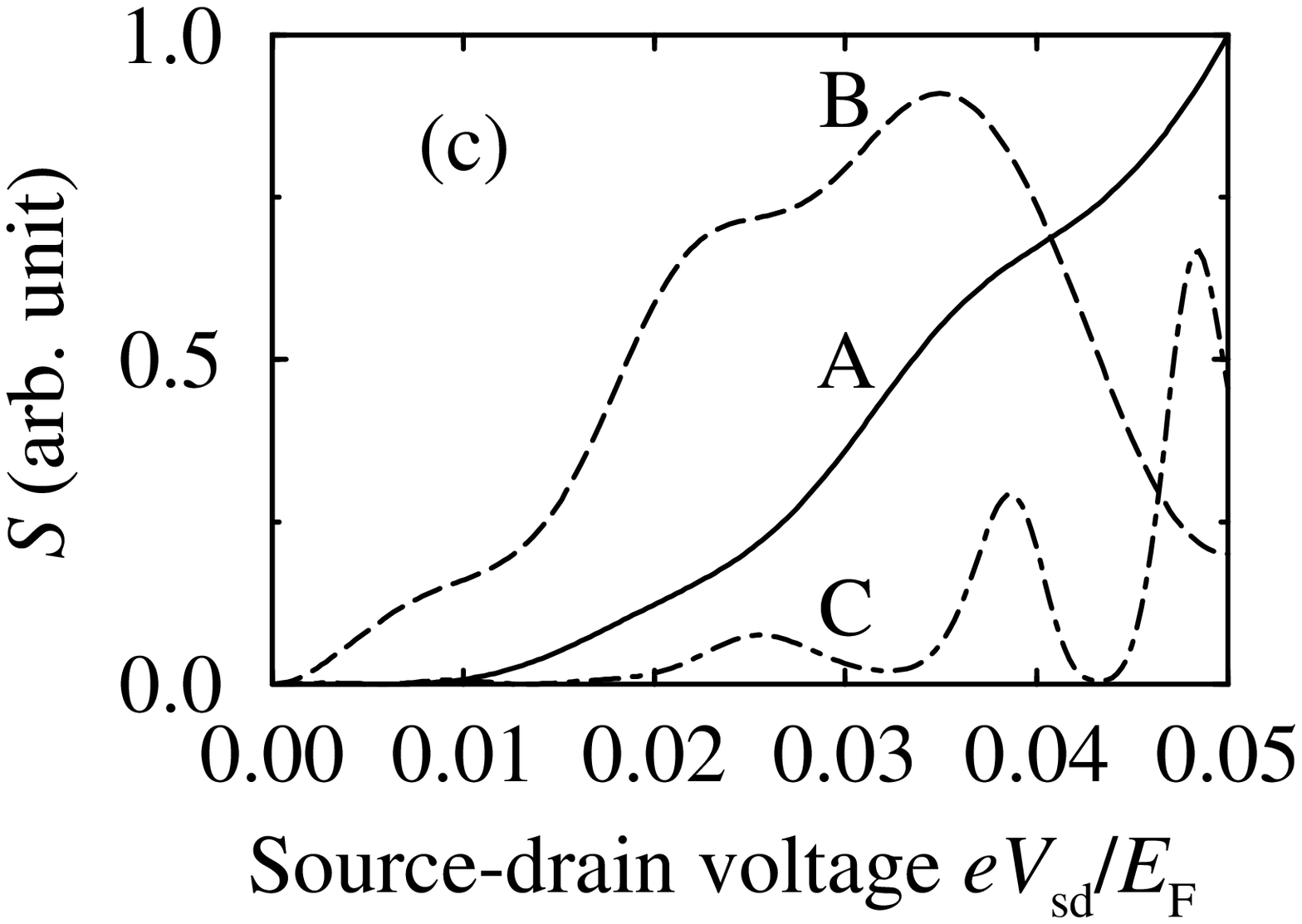,width=9cm}}
\vspace*{5 mm}
\caption{
Interference part of the current (panel $a$), transconductance (panel $b$)
and noise (panel $c$) for a channel of constant width.
They are here shown as functions of the source-drain
voltage parameter $eV_{sd}/2E_{\text F}$
for 3 different values of the dimensionless gate voltage $u$
($V_g=u E_{\text F}$).
In $A$, $u=-0.03$
while in $B$, $u=0$ and in $C$, $u=0.03$.
The lengths of the channels are in all cases 
$79.6\lambda_{\text F}$,
where $\lambda_{\text F}$ is the Fermi wavelength.
The minimum width is chosen so
that $10$ modes propagate through the contact.
The hopping distance $l$ for the
fluctuator was taken to be 
$0.796\lambda_{\text F}$.
The calculation was done for the temperature $T=10^{-3}T_{\text F}$,
$T_{\text F}$ being the Fermi temperature.}
\end{figure}

In the final set of graphs, Fig.~2 (panels $a$, $b$, and $c$),
we have studied how the current, transconductance and noise vary as
functions of the applied source-drain voltage for the case of $10$ propagating
modes. Three different values of
the dimensionless gate voltage parameter $u$ have been used, $u=-0.03 \ (A), \
0.0 \ (B),\  0.03 \ (C)$.
The range for which we are on
the same conductance step in gate voltage, is approximately $-0.08<u<0.08$. The
spread in the values of the transconductance for different values of $u$ is
large
for low-, and small and almost constant for high source-drain voltages. The
reason
for this is that, when the source-drain voltage is high,
there is essentially an averaging over all interference
patterns in the energy integrations above.
In the absence of any bias at all, $V_{sd}=0$, the current is obviously zero,
but also the noise (panel $c$) is zero. The expression for the noise
(\ref{sp2}) makes it apparent that this must always be the case.

Finally, we have found no simple relation between the magnitude of the noise
and the position on the conductance plateau.
The graphs for different values
of $u$ may cross one another as is seen in panel $c$.

\section{Discussion and conclusions } \label{discussion}

Let  us  start  the  discussion  by reviewing  the  results  for  a  QPC with
two
current-carrying modes,
Eq. (\ref{sp2})-(\ref{sp2last}).
In this case we predict that
the period of noise
oscillations as a function of external parameters should be  half  the  period
of
oscillations in the current.  Verification  of  this  statement  is
important in order to confirm (or falsify) our model.

In the general case when there is no restriction on the
number of conducting modes we conclude the following.
\begin{itemize}
\item
The dependence on temperature of  the
telegraph   noise   is   determined   by  the  factor  $\cosh^{-2}$  ($\Delta
/k_{\text{B}}T$). Comparing  with   experiments one  can
determine the interlevel spacing, $\Delta$, of the elementary fluctuator (EF).
This would help to identify the nature of the EF.
\item
The  width, $\Gamma$, of the Lorentzian
that describes the frequency dependence of the noise, is also a function of
temperature.
In fact it reflects the most important dynamical 
property of the EF because the temperature
dependence is  different  for quantum mechanical tunneling and activation
(see Section \ref{generalnoise}).
\item
The   main   feature   of   our   model  is   that variations
in  external  parameters  (such as $V_g, \ V_{sd}$) are assumed to only
affect the phase  functions $\sigma_j(E)$ (see Eq. (\ref{Phase})).
The oscillatory behavior of the noise and the square of  derivatives
of the current (with  respect  to  any external parameter)  should  then  be
qualitatively similar. In particular, the square of the transconductance is
found  to  possess  a  striking  similarity with the noise in the numerical
simulation with five modes (see below).
\item
As functions of the source-drain voltage, any form of the  interference
current and noise might appear. It depends on the actual values of external
parameters.
Using different values of the gate voltage, even the sign of the current
and transconductance can be changed.
No simple relation between the noise and the gate voltage parameter
has been found for the idealized case we have studied. The graphs may
even cross one another. In other words, it seems to be irrelevant if
we are close to a conduction step (in gate voltage), or at the
center of a the plateau. Partly, this is a result of not letting the
gate voltage  vary enough so as to change the number
of propagating modes.

\end{itemize}

{}From the numerical analysis with more than two propagating modes
we conclude that there may be an observable correlation between
the noise and the square of the transconductance. In essence,
there is an additional phase factor $\sin(w_{ij}(E))$
(see Eq. \ref{sp1lambda})
in the noise.
This may lead
to favorable situations where  the noise is suppressed while the
transconductance
is enhanced, leading to a large signal-to-noise ratio.

Finally, we would like to stress that, in spite of the  idealized
character of the adopted model, our conclusions  seem  rather general.
Indeed,   any   kind   of time-fluctuations in the scattering phase shift
by any scatterer should lead to  a  similar  behavior.  In  that  case  the
parameter  $w_{ij}$ will have a different meaning
and could in principle be of order unity.

\section{Acknowledgments}

This work was supported by NorFA,
grant no. 93.30.155, by the Nordic Research Network on the Physics of
Nanometer Electronic Devices,
and by the Swedish Natural Science Research Council (NFR).
JPH also gratefully acknowledges the hospitality
of the Department of Physics at Oslo
University, Norway.

\end{document}